\documentclass[]{aa}
\usepackage{epsfig}

\begin{document}

   \title{The INTEGRAL IBIS/ISGRI System Point Spread Function and Source Location Accuracy\thanks{Based on
observations with INTEGRAL, an ESA project with instruments and science
data centre funded by ESA member states (especially the PI countries:
Denmark, France, Germany, Italy, Switzerland, Spain), Czech Republic and
Poland, and with the participation of Russia and the USA.}}

   \author{
A. Gros \inst{1},
A. Goldwurm \inst{1},
M. Cadolle-Bel \inst{1},
P. Goldoni \inst{1},
J. Rodriguez\inst{1,5},
L. Foschini \inst{2},
M. Del Santo  \inst{3},
P. Blay \inst{4}
}

   \offprints{A. Gros : Aleksandra.Gros@cea.fr}

   \institute{CEA Saclay, DSM/DAPNIA/SAp, F-91191 Gif sur Yvette Cedex, France
   \and IASF/CNR, sezione di Bologna, via Gobetti 101, 40129 Bologna, Italy
   \and IASF/CNR, sezione di Roma, via del Fosso del Cavaliere 100, 00133 Roma, Italy
   \and GACE, Instituto de Ciencia de los Materiales, Universidad de Valencia, P.O. Box 22085, 46071 Valencia, Spain
   \and Integral Science Data Center, Chemin d'Ecogia, 16, CH-1290 Versoix, Switzerland.
   }

\date{Received ; accepted}
\authorrunning{Gros et al.}
\titlerunning{IBIS SPSF}
\abstract{ The imager on board INTEGRAL (IBIS) presently provides
the most detailed sky images ever obtained at energies above 30 keV. 
The telescope is based on a coded aperture imaging system
which allows to obtain sky images in a large field of view 
(29$^{\circ}$~$\times$~29$^{\circ}$) with an angular resolution of
12$'$. The System Point Spread Function of the telescope
and its detailed characteristics are here described along with the
specific analysis algorithms used to derive the accurate
point-like source locations. The derived location accuracy
is studied using the first in-flight calibration data on strong
sources for the IBIS/ISGRI system.
The dependence of the calibrated location accuracy with
the signal to noise ratio of the sources is presented. 
These preliminary studies demonstrate that the IBIS/ISGRI telescope 
and the standard scientific analysis software allow source
localizations with accuracy at 90$\%$ confidence level better than 1$'$ for sources
with signal to noise ratios $>$ 30 over the whole field of view, in agreement
with the expected performances of the instrument.
}
\maketitle

\unitlength=1cm
\begin{figure*} [t]
 \begin{picture}(9,3)(0,3)
   \epsfig{figure=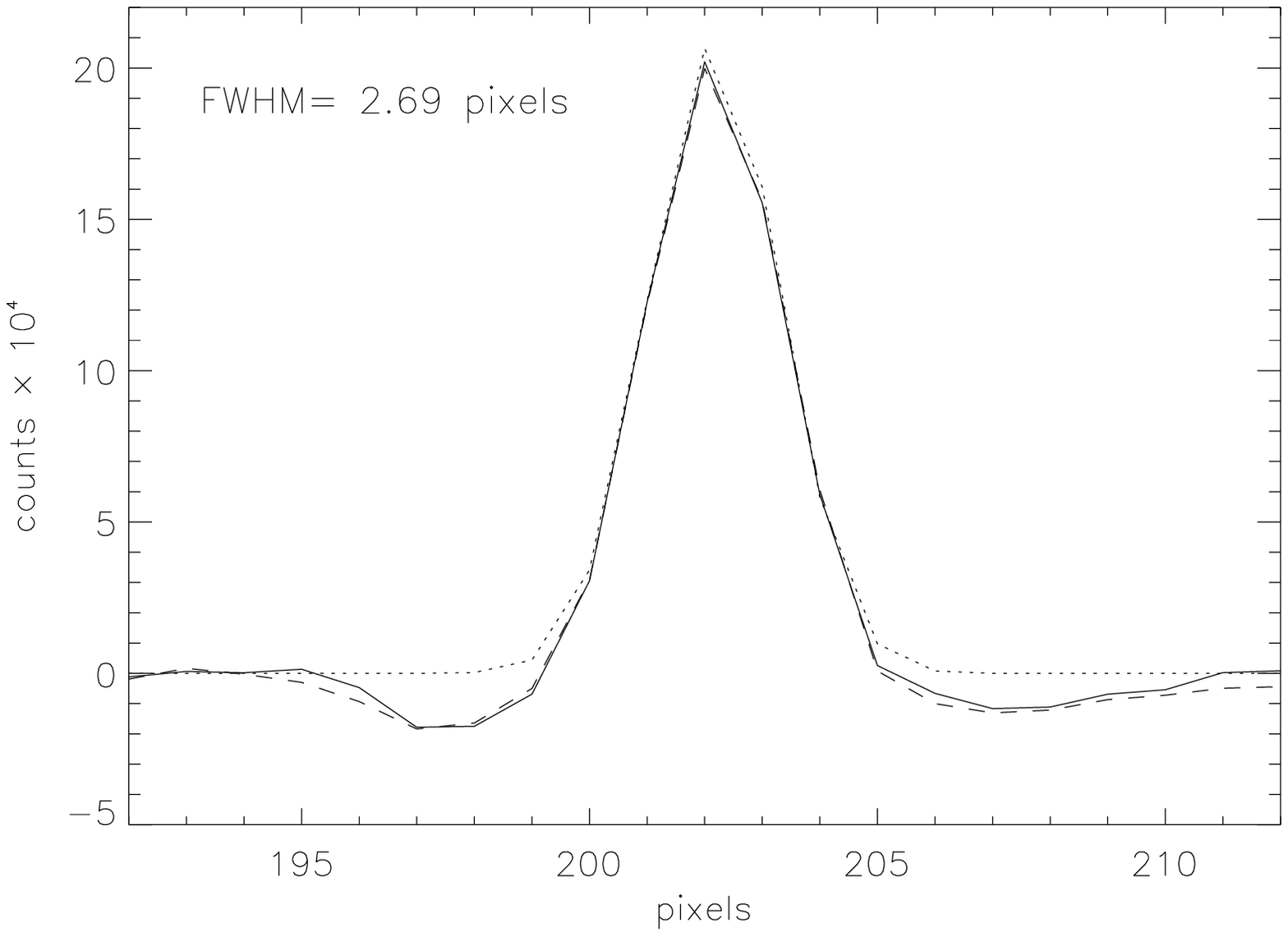,height=5cm,width=8cm}
 \end{picture}
 \begin {picture}(9,3)(0,3)
   \epsfig{figure=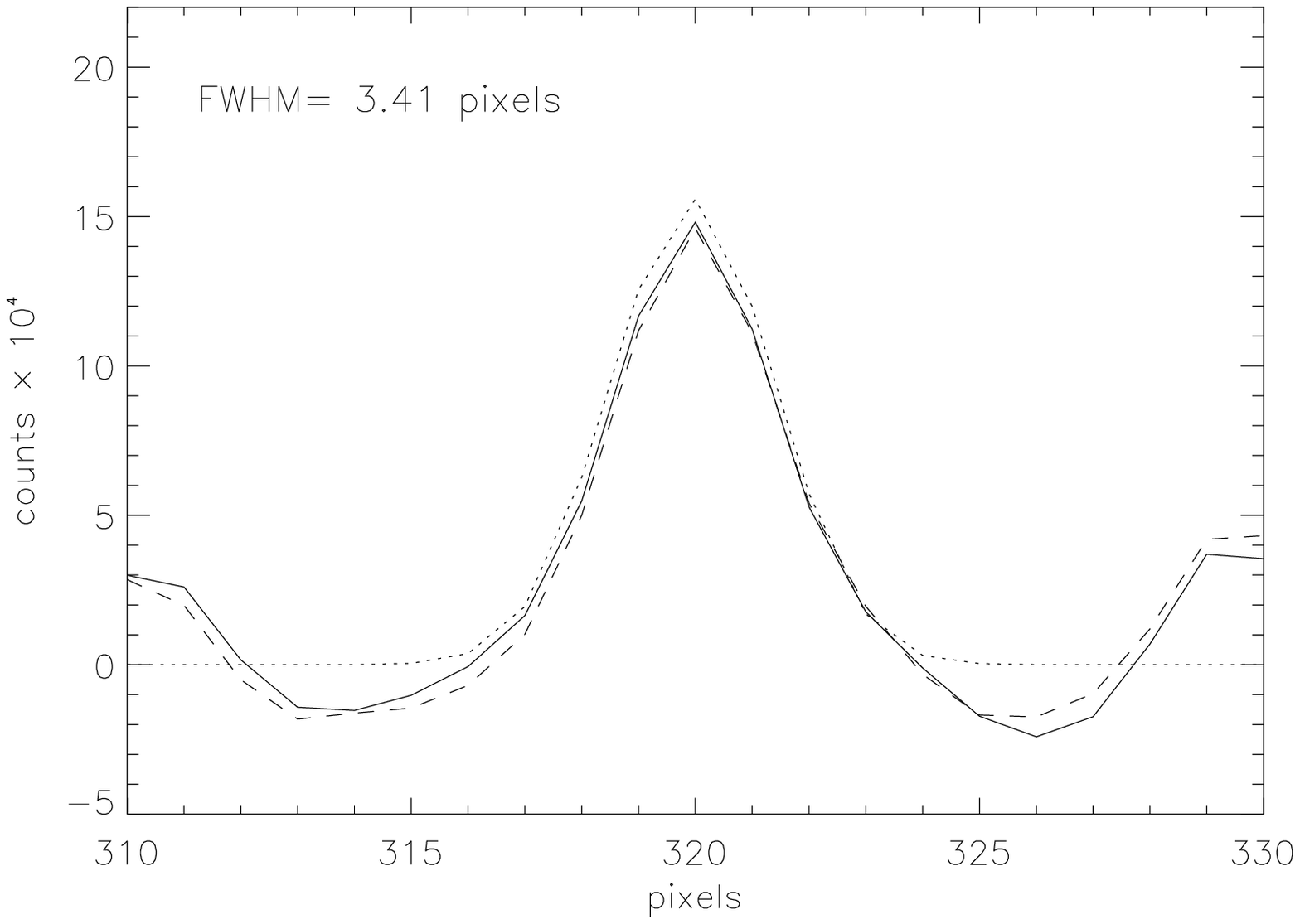,height=5cm,width=8cm}
 \end{picture}
\vspace{2.5cm}
 \caption{The ISGRI SPSF compared to
the analytical approximation used to fit source positions in the FCFOV (left)
and in the PCFOV (right). The reconstructed source profiles (solid line)
before normalization to the FCFOV for the Crab and
the fitted bidimensional Gaussians (dotted line) are shown
(fitted FWHM are 2.69 pixels for the FCFOV
and 3.41 pix for the PCFOV). Also shown is the source model
profile (broken line) computed at the position derived from the Gaussian fit.
} \label{SPSFPeak}
\end{figure*}

\section{Introduction}

The IBIS telescope (Imager on Board of the
INTEGRAL Satellite) (Ubertini et al. 2003), launched onboard the
major ESA gamma-ray space mission INTEGRAL (Winkler et al. 2003)
on October 2002, is a hard X-ray / soft $\gamma$-ray telescope
based on a coded aperture imaging system (Goldwurm et al. 2001).
The IBIS imaging system and the IBIS scientific data analysis are
described in Goldwurm et al. (2003). 
Here we discuss more in detail the specific image data analysis 
procedures used to evaluate the point source locations and 
we present the characteristics of the System Point Spread Function
obtained with the implemented analysis.  
We also provide preliminary results on the source location accuracy 
obtained from the first in-flight calibration data of ISGRI, 
the low energy (15-1000 keV) detector of IBIS (\cite{FL03}).

\section{The IBIS image decoding and the associated
System Point Spread Function}

In coded aperture telescopes (Dicke 1968, Fenimore \& Cannon 1978,
Caroli et al. 1987, Goldwurm 1995), 
the source radiation is spatially modulated by a mask of
opaque and transparent elements before being recorded on a
position sensitive detector.
Reconstruction of the sky image is generally based on a
correlation procedure between the recorded image and a decoding
array derived from the mask pattern.
For the IBIS system, where the mask is built by the replication
of a Modified Uniformly Redundant Array (MURA)
basic pattern (Gottesman $\&$ Fenimore 1989) 
of the size of the detection plane, the
raw image projected on the detector by a source in the
Fully Coded Field of View (FCFOV) (sky region from where the
recorded source radiation is fully modulated by the mask) will be
a shifted version of the mask basic pattern. A source in the
Partially Coded FOV (PCFOV) will instead project
only a part of the mask pattern. The total recorded image
(the shadowgram) is therefore the sum of the shadows projected by all
sources in the FOV plus a background term.
For those masks patterns for which a {\it correlation inverse
matrix} exists, the case of the IBIS system (Goldwurm et al. 2003), 
the sky image is reconstructed by correlation of the shadowgram 
with a decoding array obtained from its correlation inverse matrix
(Fenimore $\&$ Cannon 1981).
For those systems (and with a perfect detector) the resulting sky image 
of a single point-like source in the FCFOV will have a main peak at the
source position, flat side lobes in the FCFOV and coding noise
with 8 main source ghosts in the PCFOV.
For a PCFOV source, a main
peak will still be present at the source position but coding noise
can extend all over the image starting from the close proximity
of the main peak and ghosts will appear in the FOV.
The distribution of the coding noise does depend on the mask 
pattern used. For example URA masks built from quadratic residues
(Fenimore \& Cannon 1978)
have a high degree of symmetry along the axis of the mask, unlike those
built from Hadamard arrays (Proctor et al. 1979). 
This leads the coding noise to concentrate
along the image axis passing through the source main peak, 
producing positive and negative sidelobes (see Fig. 3 in Goldwurm et al. 2003). 

We refer to the System Point Spread Function (SPSF) as the spatial
response of the system to a point-like source {\it after} image
deconvolution, i.e. considering the decoding process
(Fenimore \& Cannon (1978). 
We will discuss here the SPSF relative to the type of deconvolution
described in Goldwurm et al. (2003), used in the
IBIS standard data analysis and optimized to obtain the best
signal to noise (S/N) for point-like sources.
This is basically a kind of balanced finely sampled cross correlation 
(Fenimore \& Cannon 1981) extended to the PCFOV 
(Goldwurm 2001, Goldwurm et al. 2003)
and in which the sampling of the decoding array
is performed in such a way to weigh the detector pixels with the fraction of 
transparent or opaque area projected by the mask elements (sect. 3). 

The algorithm implemented is fully described in sect. 2 of Goldwurm et al. (2003).
From the mask array $M$ ($=0$ for opaque elements and $=1$ 
for transparent elements)
two decoding array $G^+$ and $G^-$ are obtained projecting the
arrays $M$ and $1-M$ over a detector pixel grid and padding the zones
outside the mask with 0s. 
Then the sky image $S$ is derived from the detector image $D$ by appling 
the following operation for each $i,j$ sky pixel and where sums run over 
all $k,l$ detector pixels 
$$ 
S_{ij}=  { \sum^{ }_{kl}G^+_{i+k,j+l}W_{kl}D_{kl} } 
- B_{ij} {\sum^{ }_{kl}G^-_{i+k,j+l}W_{kl}D_{kl}}
$$
where $B_{ij} = {\sum^{ }_{kl}G^+_{i+k,j+l}W_{kl} \over \sum^{ }_{kl}G^-_{i+k,j+l}W_{kl}}$
is the balance array and the array $W$ is set to 0 for dead or noisy pixels 
and to 1 for active good pixels.
The correlation performed is balanced in the sense that in absence of sources
and with a constant background term, the reconstructed images are flat.
The decoded images are then properly renormalized to the reconstructed
counts in the FCFOV.

The above deconvolution for a perfect URA coded mask system 
would provide for a source in the FCFOV 
a SPSF close to a pyramidal function with totally flat sidelobes in the FCFOV
(Fenimore \& Cannon 1978).
However since the coding is not perfect, due to several instrumental effects 
like those produced by dead zones between pixels and 
detector modules and by the supporting structures, 
the reconstructed image will contain source sidelobes.
In the PCFOV sidelobes are inherent to the decoding process even
for a perfect detector system and the reconstructed source peak 
for this decoding procedure can be larger or
distorted because of the imperfect coding (sect. 4).
Because of the presence of the ghosts and of the coding noise,
the general sky image reconstruction process is an iterative procedure 
that combine the initial decoding process to the analysis, modelling 
and cleaning of sidelobes for each source detected in the field.
In crowded fields the initial source localization may be disturbed by the presence
of other source sidelobes, but as the modelling of the sky is improved
during the iterations the localization can improve. 
However the performances of the localization in this case depends
on the number and distribution of sources in the FOV and the performances 
of the iterative procedure including the convergence criteria used.
We present here the characteristics of the SPSF  
at the first step of the iterative procedure after the first decoding
and before the sidelobe cleaning,
and we focus on the characteristics of the main peak of the SPSF around 
the source position. 

The main peak angular width will be approximately of the projected angular 
size of the mask element, i.e. for the IBIS telescope $\approx$ 12$'$ (FWHM). 
However the exact shape of the SPSF depends also on
the detector spatial resolution, the real features of the
telescope (e.g. presence of dead and off-pixel zones, mask thickness, etc.)
and on the decoding algorithm.

\section {Analytical approximation and fitting procedure of the SPSF}

The pixels of the IBIS detectors are smaller than the mask
elements. The ratio $R$ of the linear
mask element size to the linear pixel size is 2.43 for ISGRI. 
In order to better sample the shadowgram, the corresponding
decoding array is also sampled at the pixel scale by
redistributing its values according to the fraction of pixel area
covered by a given projected mask element. The resulting decoded
image has pixels of the angular size of the detector pixels
($\approx$ 4.94$'$). This way of sampling the decoding array
before deconvolution optimizes the signal to noise ratio of
point-like sources in the reconstructed image, as it takes into
account the blurring induced by the finite spatial resolution of
the detector (discrete pixels) (Cook et al. 1984). 

For this
reconstruction the theoretical SPSF peak in the FCFOV is
space-invariant and given by the convolution of two
square-pyramidal functions with FWHM's equal to 
$w_{\rm m}$, mask element size, and $w_{\rm p}$, linear pixel size,
respectively. Using the central limit theorem we see that the
convolution of these functions can be approximated by a
bi-dimensional Gaussian with a width of $w_{\rm spsf}$ $ \approx \sqrt{
w_{\rm m}^2 + w_{\rm p}^2} $. For IBIS/ISGRI, using pixel units,
$w_{\rm p}$=1, $w_{\rm m}$=2.43, $w_{\rm spsf}$ = 2.62. 
This is slightly worse than the theoretical angular
resolution of one projected mask element $w_{\rm m}$. The standard
imaging analysis procedure of the IBIS data performs, for each
detected significant excess in the deconvolved image, a $\chi^2$
fit between an image sector around the source peak and a
bi-dimensional Gaussian. The free parameters are the centroid
position of the Gaussian (2 parameters), the 2 (variable in PCFOV) 
widths along the 2 axis, the amplitude of the Gaussian and a
constant level (background). 

In Fig.~\ref{SPSFPeak} we show the
reconstructed peak of a strong point-like source for the
IBIS/ISGRI system, versus the best fitted bi-dimensional Gaussian,
in both the FCFOV (left) and the PCFOV  (right). The source is the
well known Crab (nebula and pulsar) which was observed with
INTEGRAL in February 2003. The best fit Gaussian width is about
2.65 pixels which is compatible with $w_{\rm spsf}$. A bi-dimensional
Gaussian function is a reasonable approximation of the SPSF peak
except for the negative wings around the source, due to the non
perfect coding of the detector plane (in the FCFOV mainly due to
the dead zones). These side lobes will be corrected using the
source model once its fine position is determined. 
Fig.~\ref{SPSFPeak} shows also the shape of the peak obtained from the
deconvolution of the source model computed {\it a posteriori} for
the position obtained with the Gaussian fit. One can see that the
SPSF shape (including the wings) is well reproduced by the model. 
However, the
computation of such a model and its deconvolution is time
consuming and thus difficult to use for fine determination of the
source position. The source location is therefore determined by a
fitting procedure with a Gaussian function and the errors by the
standard computation involving the curvature matrix  (Press et al.
1996). The error computation assumes gaussian distribution of the
counts and also independence between pixels. In such decoded image
the sky pixels are instead highly correlated in particular on
length scales of the mask elements. Moreover the residual
background structures may make the distribution highly
non-gaussian. While the optimization procedure is still valid, the
derived goodness of the fit and error determination may suffer of
these conditions. The computed formal statistical error can be
underestimated and systematic errors may be dominant,  therefore
it is important to evaluate the uncertainties using in-flight data
and studying the systematic effects in the procedure.

\section{The shape of the SPSF in the PCFOV}

\begin{figure}
\centering {\epsfig{file=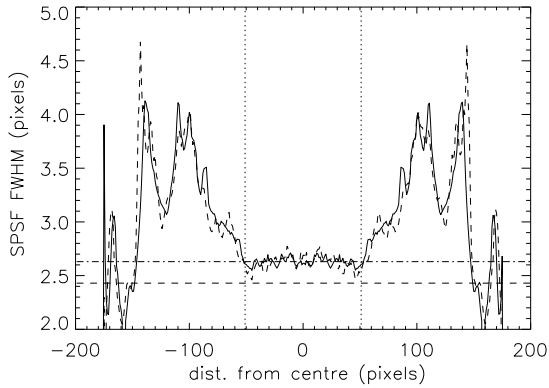 ,width=80mm} }
\caption{The variation of the width of the SPSF (for the adopted 
decoding and before sidelobe cleaning) along the FOV of
the IBIS/ISGRI telescope, for the two axis (solid line for the Y
axis and broken line for the Z axis). The two horizontal 
lines indicate the widths $w_{\rm m}$ and $w_{\rm spsf}$ and the two dotted
vertical lines the limits of the FCFOV.} \label{fig:Width}
\end{figure}

Outside the FCFOV the SPSF is not space invariant even if the
detector plane were perfect, because the optimum properties of the
MURA mask are not respected. 
The shape of the SPSF we obtain for the decoding process applied can be
distorted or enlarged (Fig.~\ref{SPSFPeak}, right).
We have studied the properties of the SPSF in the
total FOV of the telescope and the performances and limitations of
the Gaussian fitting procedure using the source model.
Fig.~\ref{fig:Width} shows the fitted width of the SPSF
on both axis for source positions all along the FOV. The
width is approximately constant in the FCFOV and ranges 
around 2.6. In the PCFOV
however it increases  up to values of about 4 pixels ($\approx$
20$'$). The behavior is symmetric with respect to the image center
and similar along the two axis. As a consequence off-axis sources 
can appear elongated or spread in the reconstructed images.

By studying the offset between the fitted and input source 
model position, we have derived a measure of the systematic bias
introduced by the approximation made for the form of the SPSF. 
The effect is not large ($<$~0.5$'$) and can be estimated. An
automatic correction of the bias has been included in the 
localization procedure.
From the first analysis of in flight IBIS/ISGRI data, a systematic offset
between the IBIS telescope axis and the axis of the satellite star
sensors used to reconstruct the absolute spacecraft attitude was
found. This predominant ($\approx$ 8$'$-10$'$) effect, was
measured using the localization procedure described here and the
available INTEGRAL data. A matrix for misalignement correction was
computed (Walter et al. 2003) and included in the localization
algorithms which operate the conversion between sky
pixels and celestial coordinates. The imaging software has already
provided good results for the source absolute positioning with
IBIS/ISGRI data (see this volume for a number of results).

In the following we characterize the performance of the system
(the telescope and the software) in term of the point source
location accuracy obtained with the most recent version of the
IBIS/ISGRI specific analysis software that will be implemented in
the October 2003 release of the ISDC Off-line Scientific Analysis 
(OSA 3.0). The location algorithms have been improved and
provide now location accuracies for off-axis sources comparable to
those of on-axis sources, therefore with better results than 
previously reported (Walter et al. 2003).

\section{The IBIS point source location error}

\unitlength=1cm
\begin{figure*} [t]
 \begin{picture}(9,3)(0,3)
   \epsfig{figure=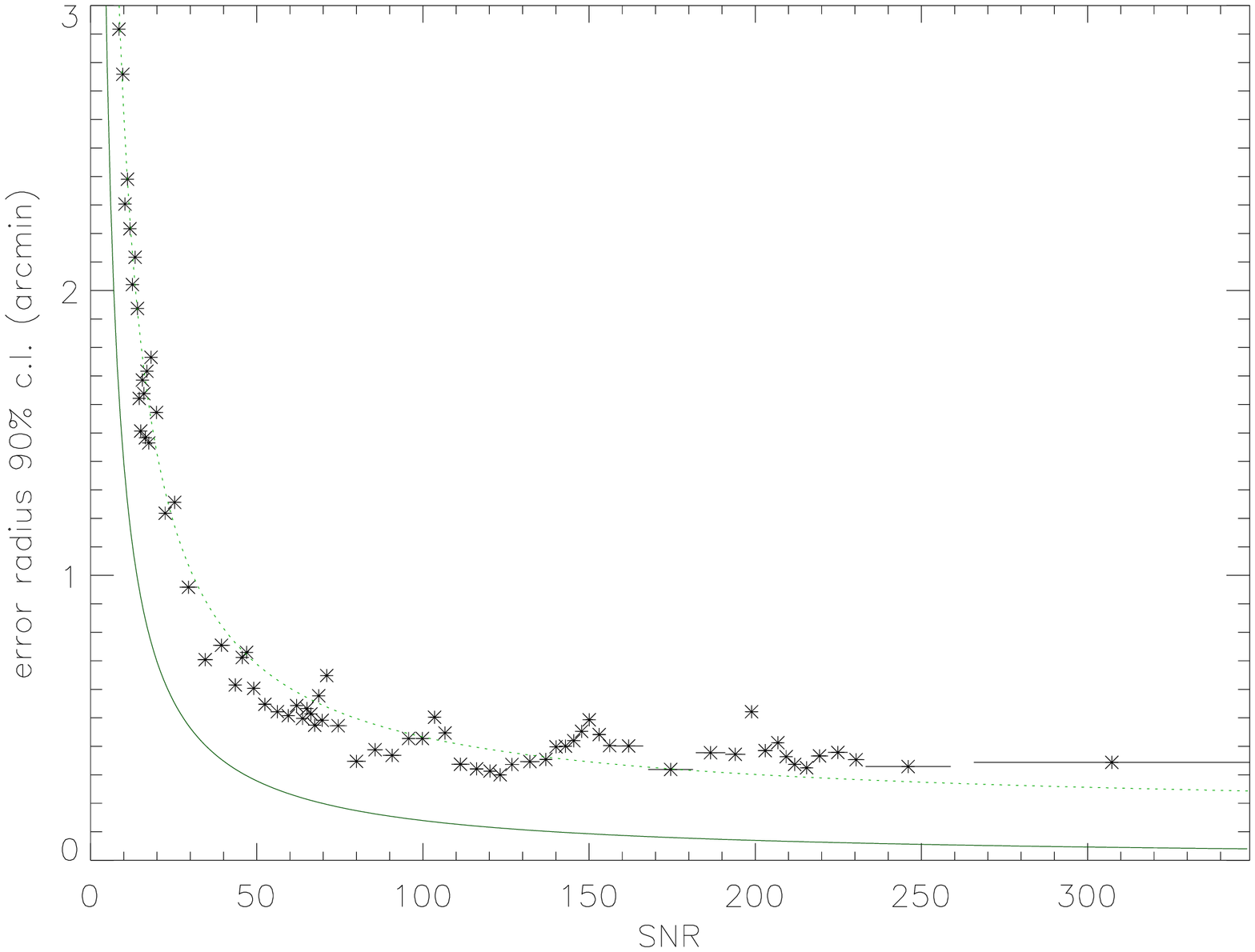,height=9cm,width=6.5cm,angle=90}
 \end{picture}
 \begin {picture}(9,3)(0,3)
   \epsfig{figure=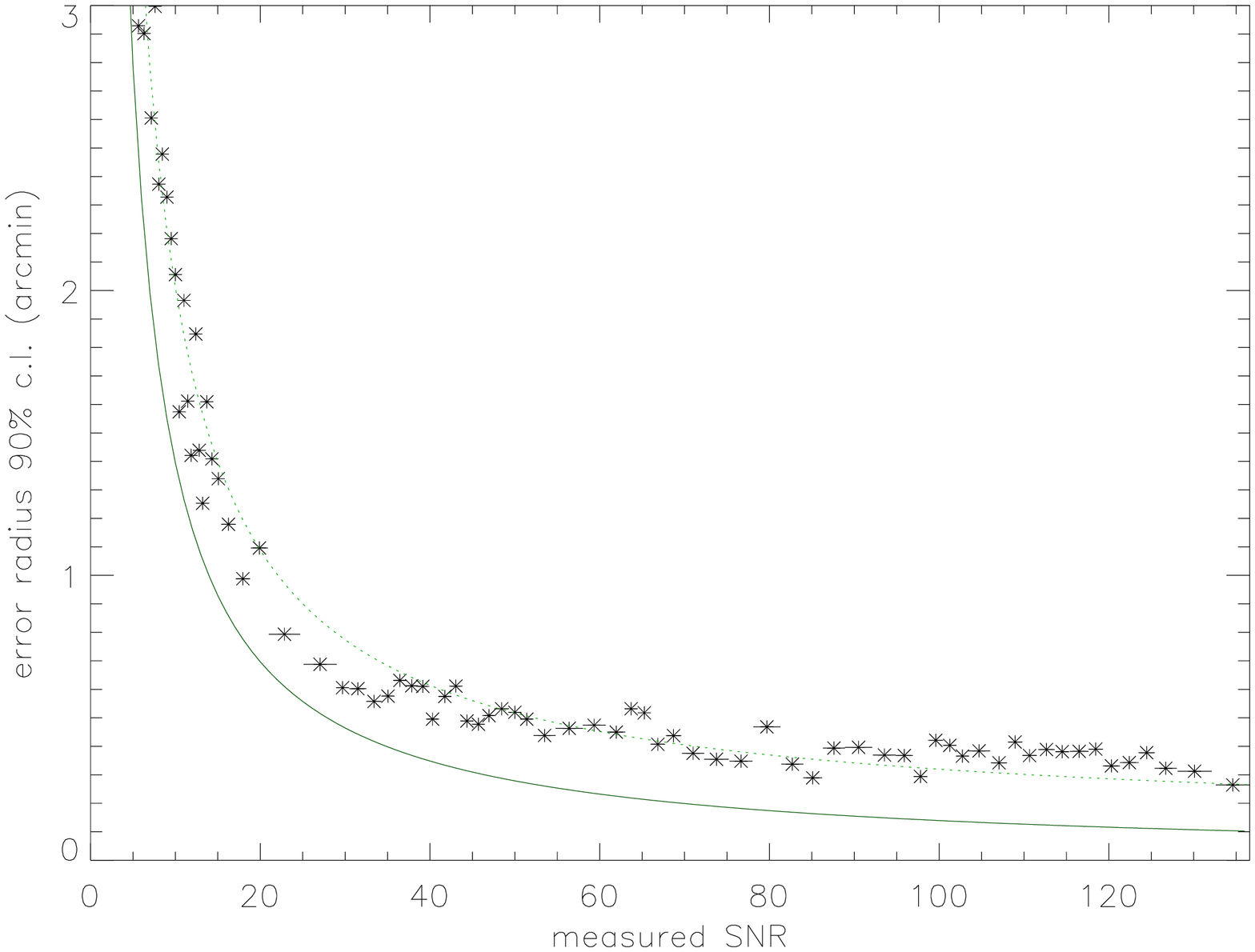,height=9cm,width=6.5cm,angle=90}
 \end{picture}
\vspace{2.5cm} \caption{The 90 $\%$ confidence level (c.l.) 
error radius on the point-like source location of the IBIS/ISGRI 
telescope vs. the source signal to noise ratio (S/N), as
derived from in-flight calibrations. 
S/N bins include 30 measures of the offset.  
The values are compared to the theoretical 90$\%$ c.l. PSLE (solid line). 
The plotted data were fitted with a function of the type 
$y = {a x^{c}} + b$ (broken line). {\it Left:} values derived from $\sim 2000$ 
offset measures of reconstructed Crab, Cyg X-1 and Cyg X-3 locations 
in energy bands between 20 and 300 keV, 
with source positions ranging between 0$^{\circ}$ and 14$^{\circ}$ from the telescope axis.
The best fit parameters are $a~=~22.1$, $b~=~0.16$ and $c~=-0.95$.
{\it Right:} Same values plotted versus a {\it measured} S/N (see text).} 
\label{LESN}
\end{figure*}

From the formal errors computed through the curvature matrix, it
can be shown that the average Point Source Location (statistical)
Error (PSLE) for an optimum coded aperture system with a defined
SPSF depends on the source signal to noise
ratio (S/N) as $ PSLE~\div~{1 \over R~\times~(S/N) }$.
The S/N is the ratio between the reconstructed source peak and
the computed statistical standard deviation associated
to the peak pixel.
Using computation of such error and simulations performed on a
{\it perfect system} with the geometrical characteristics of the
IBIS/ISGRI telescope, we derived the expected PSLE dependence on
the S/N (Goldwurm et al. 2001). This curve represents the best
theoretical accuracy of point-like source location which can be
attained by such a system.

With the described analysis software we have derived the locations
of several strong known sources, for different pointings and
energy bands. This has provided us with a number of measured
source locations in a wide range of S/N ratios and angles from the
telescope axis. We have computed the offsets between the derived
and catalogue source positions and compared their 90$\%$
confidence level dispersion to the theoretical PSLE values. 
Fig.~\ref{LESN}(left) shows the results obtained combining all
measures performed for source S/N $\geq 6$. The S/N bin widths
were defined to include a constant number of
measured offset values, in order to have similar precision in each
bin. Sources at any distance from the telescope axis were measured
and no dependence with axis angle or energy is observed. 
It can be seen that measured offsets are typically comprised between 
$3'$ and $20''$ and are better than $1'$ for S/N~$>$~30.
Although the points do not exactly follow the theoretical PSLE
curve, there is a clear trend compatible with the expected
dependence on S/N (see derived parameters in Fig. \ref{LESN}). 
At high S/N the dispersion reaches a constant level of about 20$"$
which shows the maximal accuracy obtained. Residual
systematic effects (for example due to background structures) 
may still influence the dispersion, and
work is in progress to fully evaluate their impact on the
source location determination.

In Fig.~\ref{LESN} (right) we
report the error radius versus a {\it measured} S/N rather 
than the pure statistical S/N. The measured S/N is obtained using
a standard deviation computed directly from the reconstructed sky image. 
In our analysis we measured the standard deviation 
in a region of 2.5$^{\circ}$~$\times$~2.5$^{\circ}$ around the 
source after cleaning the side lobes of all sources detected in the
FOV and neglecting the source peak regions. This way to evaluate the S/N
includes an estimate of the residual systematic errors. 
The derived offsets are now much closer to the theoretical PSLE curve
and indicate that an even better evaluation of location error may
be obtained when systematic effects will be fully accounted for.

We stress that these results involve fields with few detected
sources. For crowded fields the location
estimate may depend on the efficiency of the iterative cleaning algorithm
and a full detailed evaluation is not yet available. On the other hand,
we have verified that the location accuracy obtained by applying
the same algorithm and approximation of the SPSF to mosaicked
images (Goldwurm et al. 2003) is further improved. The results we
have obtained for selected sources in combined images show that
their localization indeed improves for a given statistical S/N. 
This may be due to the fact that the systematic effects present in single
pointings are smoothed when images are summed. Using Cyg X-1 data,
offsets of the order of $5''$-$10''$ have been obtained for
S/N values $>$ 300.

These results show that the IBIS/ISGRI telescope coupled to the
analysis procedures we have developed provide
point-like source locations with accuracies which fully comply
with the expected performance of the instrument.

\begin{acknowledgements}
A.~Gros and J.~R. acknowledge financial support from the
French Space Agency (CNES). L.~F. and M.~D.~S. acknowledge
financial support from the Italian Space Agency (ASI) and
the hospitality of the ISDC.
\end{acknowledgements}

\end{document}